\documentclass[10pt,aps,prl,twocolumn,superscriptaddress]{revtex4-2}
\usepackage[colorlinks,allcolors=blue]{hyperref}
\usepackage{amsmath,amsfonts,tikz}
\usepackage{graphicx,xcolor,multirow}
\usepackage{mathrsfs}

\def\bk{{\mathbf{k}}}

\begin{document}
\title{Defining an universal ``sign" to strictly probe phase transition}
\author{Nvsen Ma}
\affiliation{School of Physics, Beihang University, Beijing 100191, China}

\author{Jun-Song Sun}
\affiliation{School of Physics, Beihang University, Beijing 100191, China}

\author{Gaopei Pan}
\email{gppan@iphy.ac.cn}
\affiliation{Beijing National Laboratory for Condensed Matter Physics and Institute of Physics, Chinese Academy of Sciences, Beijing 100190, China}
\affiliation{School of Physical Sciences, University of Chinese Academy of Sciences, Beijing 100190, China}

\author{Chen Cheng}
\email{chengchen@lzu.edu.cn}
\affiliation{Lanzhou Center for Theoretical Physics $\&$ Key Laboratory of Theoretical Physics of Gansu Province, Lanzhou University, Lanzhou, Gansu 730000, China}

\author{Zheng Yan}
\email{zhengyan@westlake.edu.cn}
\affiliation{Department of Physics, School of Science and Research Center for Industries of the Future, Westlake University, Hangzhou 310030,  China}
\affiliation{Institute of Natural Sciences, Westlake Institute for Advanced Study, Hangzhou 310024, China}
\affiliation{Lanzhou Center for Theoretical Physics $\&$ Key Laboratory of Theoretical Physics of Gansu Province, Lanzhou University, Lanzhou, Gansu 730000, China}

\begin{abstract}
The mystery of the infamous sign problem in quantum Monte Carlo (QMC) simulations mightily restricts applications of the method in fermionic and frustrated systems. A recent work [Science \textbf{375}, 418 (2022)] made a remarkable breakthrough in the sign problem by pointing out that the sign can be used to probe phase transition. In this work, we proposed a general argument based on the definition of the sign that is related to the difference in free energy between the original and reference systems to clarify that the sign problem and phase transition cannot always be strictly related. The sign can exactly probe phase transition only if the free energy in the reference system is flat under variable parameters, which is almost impossible to design. Generally speaking, the conclusion that the sign can probe phase transition is survivorship bias without universality. To solve this problem, we define a modified sign that excludes the influence of the reference system, which can probe the phase transition strictly. The work gives an unbiased solution for detecting phase transition by the new modified sign.
\end{abstract}

\maketitle
\section{Introduction} 
%It is well known that studying a quantum many-body problem via analytical ways can be extremely difficult or even impossible. Thus, in order to understand the enormous interest phenomena with rich physics in those systems, different numerical approaches have been generated and developed rapidly in the last few decades, among which the Quantum Monte Carlo (QMC) is one of the most effective and widely used methods. 
In a quantum many-body system, the Hilbert space grows exponentially as the system size increases, and quantum Monte Carlo (QMC)  tries to overcome the exponential-wall difficulty through importance sampling and focuses on a fraction of Hilbert space. Compared with other extensively used computing methods, such as tensor network or density matrix renormalization group, QMC is especially efficient in studying models in high dimensions~\cite{Ceperley_rmp1995,Foulkes_RMP2001,Sandvik2010,Carlson_rmp2015,Assaad_book2008,Yan2019,ZY2020improved}.

However, a fundamental deficiency exists in QMC approaches, which is the notorious sign problem in many widely interested quantum systems\cite{Sugar1990exp,takasu1986monte,hatano1992representation,PhysRevB.92.045110,pan2022sign}. Some of the sampling weights in the QMC procedure can be negative or even complex in those models, which does not make sense since a distribution probability must be zero or positive. Negative weights are commonly found in fermionic or frustrated boson systems, where QMC can hardly provide meaningful results. 
To study systems with the sign problem in high dimensions and large sizes, many efforts have been made to neutralize the impact of the negative weights in the QMC
~\cite{2015BasisChange,2021BasisChange2,2020BasisChange3,2020BasisChange4,2020BasisChange5,2019BasisChange6,2020BasisChange7,Rossi2017determinant,Rossi2017polynomial,Wessel2017,DEmidio2020}.
These approaches are at least partly model-dependent and cannot conveniently be extended to general situations. Moreover, many works show that some interacting models may have intrinsic sign problems that a local unitary transformation cannot overcome~\cite{hastings2016quantum,ringel2017quantized,Intrinsic1,Intrinsic2}.

Most recently, another viewpoint is that one can use the sign problem rather than avoid it. While there has been numerical evidence indicating the severity of the sign may be related to the phase where the QMC simulation takes place~\cite{White1989,Mondaini2012,Kung2016,Wessel2017,Gotz2022}, the authors of Ref.~\cite{2021QPT} systematically investigate the sign in several interacting models and suggest a close relationship between the average sign of the sampling weight and phase transitions. In their work, the simulation results in some models show that the average sign value decreases to a minimum where the quantum phase transition (QPT) occurs, while in the other models, the derivative of the sign value reaches an extremum at the phase transition point (PTP). Thus, %the phase diagram information can be extracted from QMC simulation in those systems with a sign problem. 
the phase transition points can be extracted with the help of the average sign of the sampling weight in Ref.~\cite{2021QPT}, which could help solve the sign problem in QMC if one can easily generalize the conclusion to other spin-frustrated or fermionic systems.   % Namely, it seems that the sign problem has been solved in another way if the conclusion is universal. 
Then the key questions should be whether the PTP obtained from the average sign is correct and how universal the conclusion is. 

In this paper, we try to answer those questions through both computing results of particular systems and a general analysis independent of models. We do find one more fermionic system where the average value of the sign reaches its minimum close to PTP. However, further analysis through the essence of the sign unveils the uncertainly complex relationship between the sign value and PTP, which implies that the relationship found before can not be universal. After that, more numerical results are also given using QMC and the exact diagonalization(ED) simulations where the sign fails to detect a critical point. In this way, we propose a modified sign directly related to the free energy of the computing system. With a carefully designed computing algorithm, the modified sign probes a phase transition in one frustrated spin system at the end.  % and its computing algorithm at the end of our work. which successfully probes a phase transition in one frustrated spin system.  

%On the other hand, in previous studies, the phase transition can be related to the sign value itself or its derivative in different circumstances without an explicit criterion~\cite{2021QPT,Tarat2022,Mou2022,Mondaini2022,Gotz2022,Ding2022}. This in itself is an interesting paradox lacking theoretical explanations. This non-conformity is discussed in the present work. In principle, we have given the establishment condition when the sign can be used to probe the phase transition by the sign value itself or its derivative, although it is not an easy thing. Finally, a modified sign and its nontrivial algorithm have been proposed in this article which can strictly probe the phase transition.
\begin{figure}[!t]
    \includegraphics[width=0.7\columnwidth]{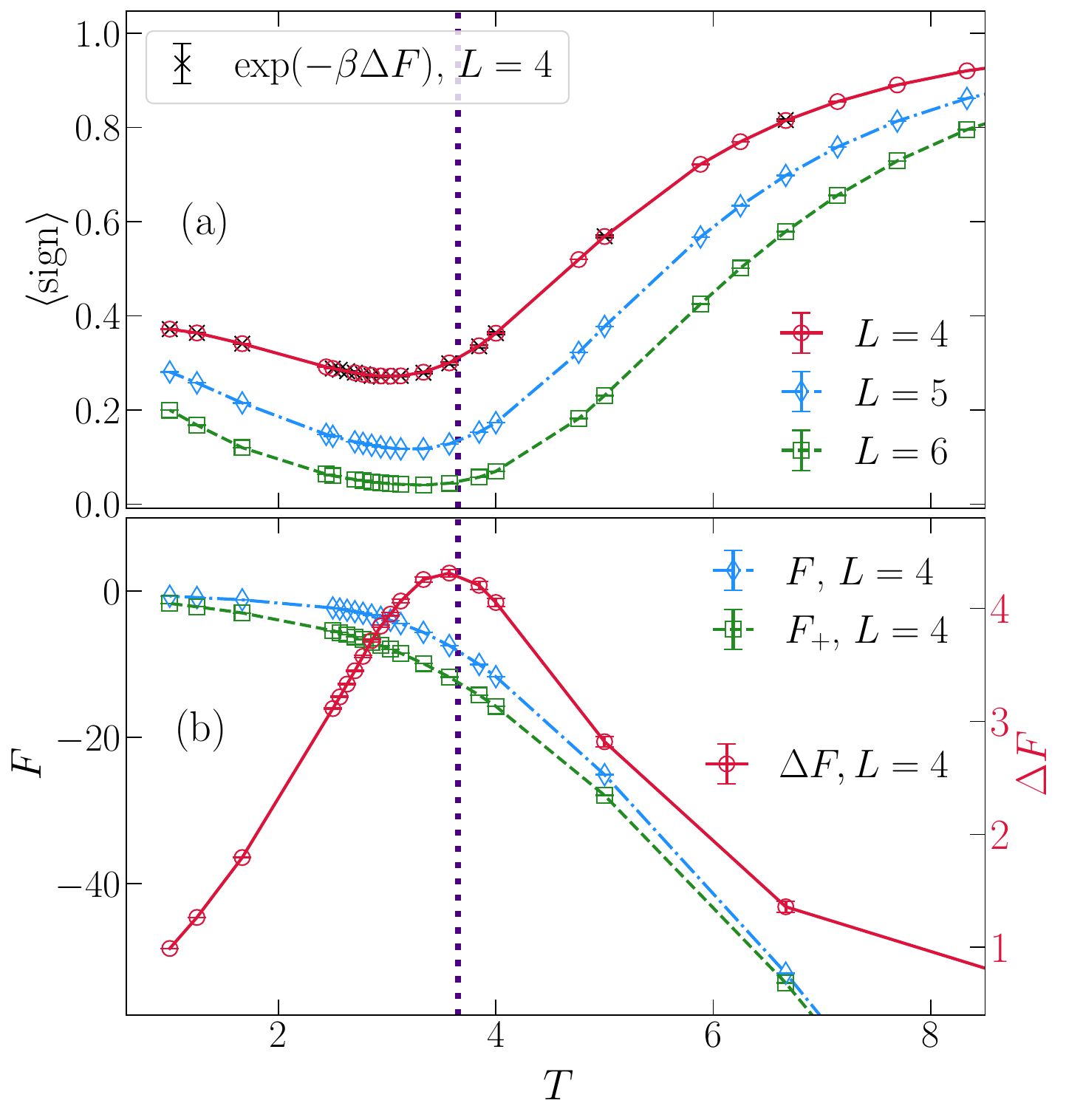}
    \caption{(a) The average sign as a function of the temperature $T$ for different system sizes $L=4,5,6$. The purple dashed line denotes the thermodynamic phase transition temperature $T_c$, which is close to the peak of the average sign. The black cross denotes $\exp(-\beta \Delta F)$, which is equal to the average sign with the same parameters. (b) The free energy of the system $Z$ and the reference system $Z_{+}$, and the difference between them. The maximum of $\Delta F$ in (b) and the minimum of $\langle \mathrm{sign} \rangle$ appear roughly at the same position. }
    \label{fig:FBLSC}
\end{figure}

\section{Positive examples by QMC}
In QMC simulation the average sign of sampling weights $\langle \mathrm{sign} \rangle$  can be computed through

\begin{equation}
\langle \mathrm{sign} \rangle=\frac{Z}{Z_+}=\frac{\sum_C W(C)}{\sum_C |W(C)|},
\label{eq1}
\end{equation}

where the partition function $Z=\exp(-\beta H)$ is the sum of simulated configurations $C$ with weight $W(C)$, and for $Z_+$, the configuration weights are all positive as $|W(C)|$, which can be considered as the partition function of a sign-free auxiliary system. In the following, the sign-free system is called the reference system.

We first calculate $\langle \mathrm{sign} \rangle$  %perform the Determinant Quantum Monte Carlo(DQMC)~\cite{Assaad_book2008} simulation
on the flat-band interaction model\cite{pan2022thermodynamic,bernevig2020tbg3,hofmannFermionic2022} with long-ranged single-gate-screened Coulomb interaction (FBLSC) using Determinant Quantum Monte Carlo(DQMC)~\cite{Assaad_book2008} .
A phase transition from the quantum anomalous Hall (QAH) insulator to the metallic state has been observed in FBLSC recently \cite{pan2022thermodynamic}. %Despite negative weights in QMC sampling, $\langle \mathrm{sign} \rangle$ decays algebraically with system size $L$ at low temperature according to the sign bound theory\cite{PhysRevB.106.035121}, which makes it possible to perform QMC simulations and probe the relationship between $\langle \mathrm{sign} \rangle$ and critical points.  %More details about the model and corresponding computational approaches are provided in Sec. I of the Supplemental Material (SM).
%Here, we study the behaviors of sign problem in the flat-band model with long-ranged Coulomb interaction, which can be used to investigate the thermodynamic properties\cite{pan2022thermodynamic} of correlated flat-band twisted bilayer graphene (TBG) at 3/4 filling.
 Even though there exists a sign problem in FBLSC, the sign bound theory\cite{PhysRevB.106.035121} makes it possible to simulate at the finite size and finite temperature, which makes it a good example to detect the relationship between $\langle \mathrm{sign} \rangle$ and PTP. 

The Hamiltonian of FBLSC in the momentum space is:
$H_I=\frac{1}{2 \Omega} \sum_{\mathbf{G}} \sum_{\mathbf{q} \in m B Z}V(\mathbf{q}+\mathbf{G}) \delta \rho_{\mathbf{q}+\mathbf{G}} \delta \rho_{-\mathbf{q}-\mathbf{G}}$ 
with $\delta \rho_{\mathbf{q}+\mathbf{G}}=\sum_{\mathbf{k}, m} \lambda_{m}(\mathbf{k}, \mathbf{k}+\mathbf{q}+\mathbf{G})\left(d_{\mathbf{k}, m}^{\dagger} d_{\mathbf{k}+\mathbf{q}, m}-\frac{1}{2} \delta_{\mathbf{q}, 0} \right)$, mBZ the  moir\'e Brilliouin zone and $m=\pm 1$ denotes the Chern band indices since we have projected the system onto flat bands. In the simulation eigenstate $\left|u_{\boldsymbol{k}, m}\right\rangle$ of Bistritzer-MacDonald (BM) model is chosen \cite{bistritzerMoire2011,bernevig2020tbg3,YiZhang2020} with $d_{\boldsymbol{k}, m}^{\dagger}$ to be its creation operator and form factor is defined as 
$\lambda_{m}(\mathbf{k}, \mathbf{\bk}+\mathbf{q}+\mathbf{G}) \equiv\left\langle u_{\bk, m} | u_{\bk+\mathbf{q}+\mathbf{G}, m}\right\rangle$.
The area $\Omega=N_k \frac{\sqrt{3}}{2} L_{M}^{2}$ with $N_k=L\times L$ is the number of $\mathbf{k}$ points in $\mathrm{mBZ}$. And $V(\mathbf{q})=\frac{e^2}{4 \pi \varepsilon} \int d^2 \mathbf{r}\left(\frac{1}{\mathbf{r}}-\frac{1}{\sqrt{\mathbf{r}^2+d^2}}\right) e^{i \mathbf{q} \cdot \mathbf{r}}=\frac{e^2}{2 \varepsilon} \frac{1}{q}\left(1-e^{-q d}\right)$ is the long-ranged single-gate-screened Coulomb interaction\cite{liuNematic2021} with $\frac{d}{2}$ the distance between graphene layer and single gate while $\epsilon$ a dielectric constant, and in the simulation we chose $d=40$ nm and $\varepsilon=7 \varepsilon_0$.

The numerical results displayed in Fig.~\ref{fig:FBLSC}(a) show that $\langle \mathrm{sign} \rangle$ reaches its minimum close to $T_c=3.65(5)$ and the location of the minimum becomes closer to $T_{c}$ as the system size increases. Our simulations on the FBLSC model again support the argument that the  $\langle \mathrm{sign} \rangle$ can help detect the critical point, which gives us more motivation to understand the internal mechanism between the sign problem and phase transition beyond specific models.

\section{General analysis about sign}
It is known that the free energy $F-T\ln Z$ and  
%partition function $Z=\exp(-\beta F)$, 
we can rewrite Eq.~(\ref{eq1}) as 
\begin{equation}
\langle \mathrm{sign} \rangle = e ^{-\beta \Delta F},
\label{eq:sign_in_F}
\end{equation}
where $\beta=1/T$ and $\Delta F=F-F_{+}$with $F$ and $F_{+}$ the free energy of the original system and the reference system correspondingly. The phase transition always occurs with a dramatic change in free energy $F=E-TS$, which separates the internal energy($E$)-dominated ordered phase at low-T and the entropy($S$)-dominated disorder phase at high-T. The quantum fluctuation tunes the QPT results from a changeable parameter $J$ for quantum systems at zero temperature. It is easy to find that the position of the critical point $T_c$ (or $J_c$ as the quantum critical point) is only related to $F$ while $\langle \mathrm{sign} \rangle$ depends on both $F$ and $F_+$.  The relationship between $T_c$/$J_c$ and $\langle \mathrm{sign}\rangle$ can be understood through $F$ and $\Delta F$. However, the connection between  $F$ and $\Delta F$ is not invariable in different systems. For simplicity, we choose three typical ones and try to understand the finding that $\langle \mathrm{sign} \rangle$ can detect $T_c$/$J_c$ with their help.

\begin{figure}[htp]
    \includegraphics[width=1\columnwidth]{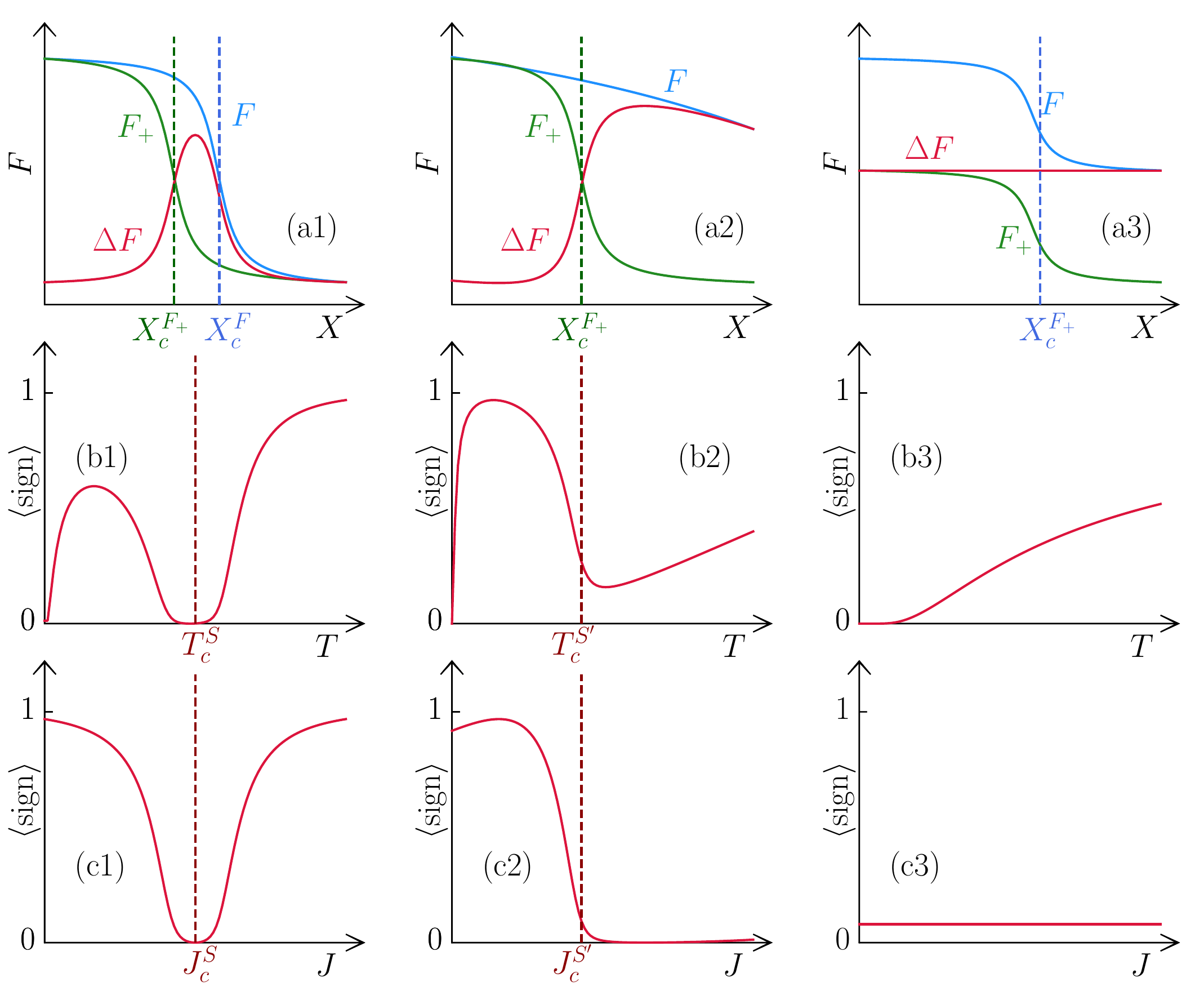}
    \caption{Schematic diagram for three typical relations between the free energy $F$ and sign, as demonstrated in three columns. The x-axis in (a1-a3) can be $T$ or $J$ when probing the finite temperature phase transition or quantum phase transition. The corresponding behaviors of sign value are displayed in (b1-b3) and (c1-c3) for the temperature-driven phase transition and quantum phase transition at a fixed temperature, respectively. Here $X_c^{F}$ ($X_c^{F_+}$) labels the critical point extracted from the rapid change of $F$ ($F_+$). In the first column, $X_c^{S}$ is extracted from the minimum of $\langle \mathrm{sign} \rangle$; in the second column, $X_c^{S}$ is extracted from the rapid change of $\langle \mathrm{sign} \rangle$.  
    }
    \label{fig:schematic}
\end{figure}

The first possible situation is that  $Z$ and  $Z_+$ are not evidently different. In this case, the $T$ (or $J$ at zero temperature) dependence of free energy in both systems is similar to each other, as illustrated in Fig.\ref{fig:schematic} (a1) and $\Delta F$ has a small maximum region covering both critical points. As a consequence, the $\langle \mathrm{sign} \rangle$ related to $\Delta F$ reach a minimum in that region as shown in Fig.\ref{fig:schematic} (b1) for the $T$-driven and Fig.\ref{fig:schematic} (c1) for the $J$-driven transition in this situation. Thus, the $\langle \mathrm{sign} \rangle$ can probe the phase transition point in the simulated system with a rough precision. 
The FBLSC model is a similar case as displayed in Fig.~\ref{fig:FBLSC}(b), which explains why the minimum of $\langle \mathrm{sign} \rangle$ can be a criterion judging the phase transition here. This is also true for the model studied
in Ref~\cite{Tarat2022}. Here and after, we neglect the trivial minimum of $\langle \mathrm{sign} \rangle$ at zero temperature resulting from the divergent $\beta$.

The second situation is shown in Fig.~\ref{fig:schematic} (a2), where the original system has no phase transition with a constantly changing free energy while $F_{+}$  drops quickly around the referenced critical point $X_{c}^{F_{+}}$. Free energies of the two systems are close to each other at low temperatures with a very small $\Delta F$ and then distinctly separate with a much larger $\Delta F$  due to the dramatic change in $F_{+}$. In this case,  $\langle \mathrm{sign} \rangle$ can also exhibit a minimum/drop $\langle \mathrm{sign} \rangle$ as Fig.~\ref{fig:schematic} (b2)/(b3), which can not be related to $X_{c}$ any more as there is no transition in the original system at all. %Thus, the minimum of $\langle \mathrm{sign} \rangle$ fails to probe the phase transition but provides misleading information. 

Besides, if the $T$ or $J$ dependence of $F$ and $F_{+}$ in Fig.\ref{fig:schematic} (a2) is interchanged with each other, there is a phase transition in the studied system then. Through previous analysis, $\Delta F$ would drop rapidly at the critical point in this situation. Therefore, the derivative of $\langle \mathrm{sign} \rangle$ rather than the sign value itself is closely related to PTP, which explains why one can use the derivative of $\langle \mathrm{sign} \rangle$ to probe phase transitions in the Ref~\cite{Tarat2022}. We should point out that the minimum of $\langle \mathrm{sign} \rangle$ or its derivative can be pretty helpful in studying phase transitions from the analysis up to now. However, the two criteria are applicable in different cases. It could not be a convincing procedure to randomly choose one of them or use both in the same system. Determining which one to choose with little knowledge of the system is also very difficult,  making extracting correct information from the sign problem hard.

For the third situation, as illustrated in Fig.\ref{fig:schematic} (a3), $F$ and $F_+$ are always far away from each other with or without phase transitions. The sign value is close to zero in a wide range of parameters in both thermal (b3) and quantum (c3) transitions. The increasing of $\langle \mathrm{sign} \rangle$ in Fig.\ref{fig:schematic} (b3) results from the decreasing of $\beta$, i.e., the increasing of temperature. In this case, no helpful information can be extracted from $\langle \mathrm{sign} \rangle$. % let alone a rough approximation of the phase transition point. 

The three examples discussed above do not cover all possibilities, but they are typical enough to show that the value of $\langle \mathrm{sign} \rangle$ are not always  as precise as expected. While $T_{c}/J_{c}$ is only related to $F$, the changing of  $\langle \mathrm{sign} \rangle$ can be influenced
by both $F$ and $F_{+}$ independent of QMC algorithm or simulated model.
%\paragraph{Exact diagonalization results.---}
%\paragraph{Example II: Failure in using sign probing phase transition.---}
\begin{figure}[htp]
    \includegraphics[width=\columnwidth]{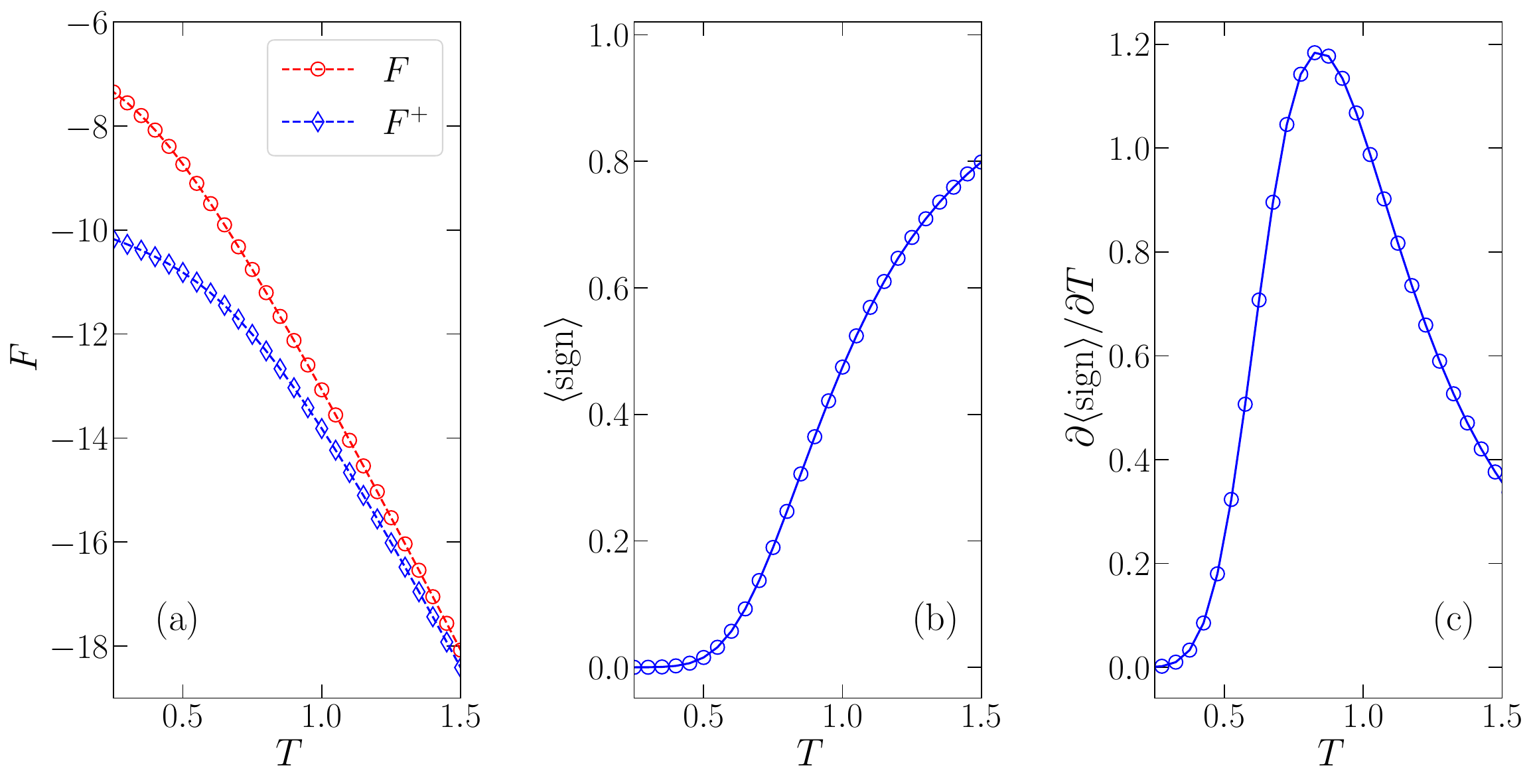}%./draft_J1J2_QPT.pdf}
    \caption{ (a) The free energy $F$ and $F_+$, as well as the their difference $\Delta F$ versus $T$. (b) The sign value as a function of $T$. (c) $\partial \langle sign\rangle/\partial T$ vs $T$. All the data are from ED calculations of the Eq.\ref{eq:Hj1j2} and Eq.\ref{eq:Hj1j2-2} on a 4$\times$4 square lattice with $J_1=1, g_1=0.7$ and $J_2=0.2, g_2=0.7$.}
    \label{fig:J1J2_FvS}
\end{figure}

\section{Negative examples} 
Still, the failure of $\langle \mathrm{sign} \rangle$ in detecting PTP can be questioned as to whether the situation in Fig.\ref{fig:schematic} (a2) and (a3) truly exists or not. This section provides practical numerical results in specific systems. Among several systems we examined, the anisotropic $J_1$-$J_2$ Heisenberg model on the square lattice can be a typical example. The $J_1$-$J_2$ Hamiltonian reads
\begin{align}
    \label{eq:Hj1j2}
    H=J_1\sum_{\langle i,j\rangle}[g_1S^z_iS^z_{j}+\frac 1 2 (S^+_iS^-_j+S^-_iS^+_j)]\nonumber\\
+J_2\sum_{\langle\langle i,j\rangle\rangle}[g_2S^z_iS^z_{j}+\frac 1 2 (S^+_iS^-_j+S^-_iS^+_j)],
\end{align}
where $\langle i,j\rangle$ and $\langle\langle i,j\rangle\rangle$ denotes the nearest and the next nearest neighbors respectively. The model in isotropic case with positive $J_1$ and $J_2$ was first introduced to describe the breakdown of N\'eel order in cuprate superconductors \cite{Inui1988,Chandra1988,Dagotto1989} and has been extensively studied in the past several decades~\cite{Singh1999,Capriotti2000,Gong2014,Yu2020,Jiang2012,Hu2013,Wang2018,Liu2018,Ferrari2020,Nomura2021,Liu2022}. It is generally accepted that the ground state phase has a N\'eel antiferromagnetic (AFM) order in the small $J=J_2/J_1$ region, and the collinear AFM order in the large $J$ region. But the physics in the intermediate region ($J_2/J_1$ around 0.5) is still unsettled, with possibilities of a gapped/gapless spin liquid phase~\cite{Jiang2012,Hu2013,Wang2018,Liu2018,Ferrari2020} or even two intermediate phases~\cite{Nomura2021,Liu2022}. 
%A temperature-driven Ising in the collinear phase has also attracted investigations in recent years~\cite{Poilblanc2021,Niggemann2021,Gauthe2022}. 
The antiferromagnetic $J_1$-$J_2$ Hamiltonian ($J_1, J_2>0$) suffers from the sign problem and leads to the failure of QMC. So here we adopt the exact diagonalization (ED) method to analyze the behavior of $\langle \mathrm{sign} \rangle$ with the help of a reference system
\begin{align}
    \label{eq:Hj1j2-2}
    H_+=J_1\sum_{\langle i,j\rangle}[g_1S^z_iS^z_{j}-\frac 1 2 (S^+_iS^-_j+S^-_iS^+_j)]\nonumber\\
+J_2\sum_{\langle\langle i,j\rangle\rangle}[g_2S^z_iS^z_{j}-\frac 1 2 (S^+_iS^-_j+S^-_iS^+_j)].
\end{align}
%The Monte Carlo method can not solve a quantum Hamiltonian with a severe sign problem. 
%To systematically investigate the relationship between the physical quantities (such as the free energy) and the sign, we present ED calculation on the systems of $4\times4$ sites with different values of $J_2$ and periodic boundary conditions. 
We calculate  $Z$ and $F$ on a diagonal basis and obtain $Z_+$ and $F_{+}$ of the reference system on the same basis. Although ED does not meet any sign problem, we can estimate the sign value $\langle$sign$\rangle=Z/Z_+$ from ED calculations of the $J_1$-$J_2$ model and its reference Hamiltonian. %The consistency between the $\langle$sign$\rangle$ from Monte Carlo calculations and its estimated value $Z/Z_+=\exp(-\beta \Delta F)$ have also been numerically confirmed in Fig.~1 in the main text.

As displayed in Fig.\ref{fig:J1J2_FvS}, the free energy $F$ of the system with Hamiltonian $H$ has no obvious mutation, while there is one in the curve of the free energy $F_+$ described by $H_{+}$, which is qualitatively the same case discussed in Fig.\ref{fig:schematic}(a2). %Although the ``mutation" is not exact because of the finite-size effect, it shows that the free energy in the reference system can affect  $\langle \mathrm{sign} \rangle$ and lead to a fake phase transition. 
One can extract a possible phase transition point %(a peak of differential function in a finite size system) 
around $T=0.7$ from the obvious change of $\langle \mathrm{sign} \rangle$ in Fig.~\ref{fig:J1J2_FvS}(b) and the peak of its differential (c), which, however, only mistakenly refers to the possible critical point of the reference system $H_+$. We want to point out that although it is very hard to distinguish a singularity and a continuous peak in finite size, it should be noticed that either can be introduced by $H_+$, which would mix up the signal of the original system. Note: According to the Mermin-Wagner theorem, we know this system has no finite temperature phase transition in fact. However, it is not important in our work because we are focusing on demonstrating the signal from the reference system can induce the sign value displaying a fake semaphore containing both a singularity (real phase transition) or a continuous peak (fake phase transition).
%The behaviors of $-\beta\Delta F$ and  $\langle \mathrm{sign} \rangle$ are consistent with the expected result in Fig.\ref{fig:schematic}(c). 
 %Here probing sign value does not provide helpful information to the controversy and interesting physics in the original model $H$.
In this example, changing $\langle \mathrm{sign} \rangle$ would bring us confusing and misleading phase transitions. Similar results are also found in some Fermion models recently~\cite{tiancheng2022,PhysRevB.107.245144}. The Ref.\cite{PhysRevB.107.245144} also found the reference system can affect the sign with a fake phase transition.
 
\section{Modified sign to probe the transition} 
The conclusion from the previous analysis is slightly disappointing, but the successful situations inspire us. The original sign is defined as $\langle sign=Z(\beta,J)/Z_+(\beta,J)\rangle$ with $\beta=1/T$ and related to both systems, which is the decisive reason for its failure. If the denominator is fixed, the sign will behave like the $Z$ only, which can strictly reflect the phase transition.
Therefore, a new kind of "sign" can be redefined as 
\begin{align}
    \label{eq:newsign}
 \langle \widetilde{sign}\rangle=Z(\beta,J)/Z_+(\beta^*,J^*),
\end{align}
where $\beta^*$ and $J^*$ are fixed constants.
\begin{figure}[htp]
    \includegraphics[width=1.0\columnwidth]{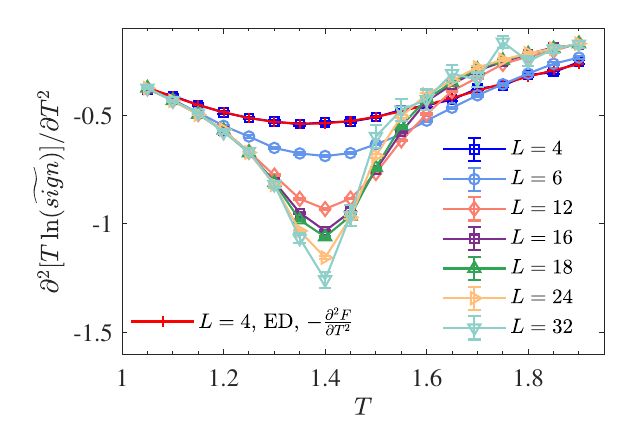}
    \caption{The temperature dependence of $\partial^2[T\ln(\langle \widetilde{sign}\rangle)]/\partial T^2$ per site using SSE on the Hamiltonian in Eq.(\ref{eq:Hj1j2}) from $L=4$ to $L=32$ on square lattice with $J_1=1$ and $g_1=g_2=-2$ . The fixed calculation parameters of the reference system when calculating $\langle \widetilde{sign}\rangle$ are: \(\beta^*=2/3\), \(J_2^*=0.2\). The computing results at different temperatures for $L=4$ agree with the second-order derivative of free energy per site of the same Hamiltonian got from ED.}
    \label{j1j2sign}
\end{figure}
In this way, the $\langle \widetilde{sign}\rangle$ is directly proportional to the $Z$, which can be calculated using QMC. %The difficulty becomes calculating the $\langle \widetilde{sign}\rangle$ in QMC. %Generally speaking, we can simulate the reference system at certain $\beta^*$ and $J_2^*$ directly using QMC with no sign problems. After that, for $Z(\beta,J)=\sum_c W_c(\beta,J)$ each configuration $W_c(\beta,J)$can be regained through $W_c(\beta^*,J^*)$ times the factor $W_c(\beta,J)/W_c(\beta^*,J^*)$ to regain the actual weight of $Z(\beta,J)$.
Taking the stochastic series expansion (SSE) QMC method as an example, the partition function can be written in the form 
\begin{align}
    \label{sse}
Z=\sum_\alpha\sum_{n=0}^\infty\frac{\beta^n}{n!}\langle\alpha|(-H)^n|\alpha\rangle,
\end{align}
with ${|\alpha\rangle}$ the expansion basis and $n$ the expansion order. In this way, 
\begin{align}
    \label{sse2}
\langle \widetilde{sign}\rangle=\langle (-1)^{n_{s}}(\frac {\beta} {\beta^*})^{n}(\frac {J} {J^*})^{n_J}\rangle,
\end{align}
where the $n_s$, $n$ and $n_J$ are the numbers of the sign operators, total operators and $J$ operators in the series~\footnote{In the definition of the original sign, the $\beta=\beta^*$ and $J=J^*$, the original sign, therefore, is $\langle (-1)^{n_{s}}\rangle$.}.

It's worth noting that $\beta$ and $J$ in Eq.~\ref{sse2} can not be far away from the $\beta^*$ and $J^*$ because the distribution for $Z(\beta,J)$ and $Z_+(\beta^*,J^*)$ should be similar. If the $\langle \widetilde{sign}\rangle$ in a wide region of $\beta$ or $J$ is necessary, which is always the case as the PTPs are usually unknown, a series of intermediate processes are also necessary. We can calculate it using
\begin{align}
    \label{sse2}
\langle \widetilde{sign}\rangle=\frac {Z(\beta,J)}{Z_+(\beta_1,J_1)}\frac {Z_+(\beta_1,J_1)}{Z_+(\beta_2,J_2)}\frac {Z_+(\beta_2,J_2)}{Z_+(\beta_3,J_3)}...\frac {Z_+(\beta_m,J_m)}{Z_+(\beta^*,J^*)},
\end{align}
in which $\beta_1, ..., \beta_m$ and $J_1, ..., J_m$ are inset points of $\beta$, $\beta^*$ and $J$, $J^*$. As the fixed $Z^+$ in $\widetilde{sign}$ can be treated as a constant $C$, the free energy of the simulated system turns to be 
\begin{align}
    \label{freenergy}
    &F=-T\ln(Z)=-T[\ln(\langle \widetilde{sign}\rangle)+\ln(C)],\\
   &\partial F/\partial T=-\partial [T\ln(\langle \widetilde{sign}\rangle)]/\partial T-\ln(C),\\
   &\partial^{2} F/\partial T^{2}=-\partial^{2} [T\ln(\langle \widetilde{sign}\rangle)]/\partial T^{2}
\end{align}
now. Thus, the free energy and its derivative of the original system can be obtained (at most with a constant) at any given temperature using the new definition of $\langle \widetilde{sign}\rangle$. 

As an example, we calculate $\langle \widetilde{sign}\rangle$ on one frustrated model in Eq.(\ref{eq:Hj1j2}) with $g_1=g_2<-1$  using SSE. The interactions of spins in the z-direction are ferromagnetic with absolute values larger than the x- and y-direction, so there should be an Ising-like phase transition at finite temperature in this system. 
%\begin{eqnarray}
%H&=J_1\sum_{\langle i,j\rangle}[-\frac 1 2 (S^+_iS^-_j+S^-_iS^+_j)+J_zS^z_iS^z_j-C_1] \nonumber\\
%&+J_2\sum_{\langle\langle i,j\rangle\rangle}[-\frac 1 2 (S^+_iS^-_j+S^-_iS^+_j)+J_zS^z_iS^z_j-C_2],
%\label{hamm}
%\end{eqnarray}
%where $J_1$ and $J_2$ stand for the antiferromagnetic nearest and second nearest couplings in the square lattice. Besides, 
%In this way, to minimize the bouncing weights using directed loop update during the SSE simulation, there should be $C_{1}=C_{2}=J_{1}/4$.\cite{pre}\zyan{Nvsen, Junsong, pls write down the correct Hamiltonian here}
The second-order derivative of free energy got from $\langle \widetilde{sign}\rangle$ presents a diverging peak with size increases as shown in Fig.\ref{j1j2sign}, which unbiasedly probes the second-order phase transition point of the model.

\section{Conclusions}
The sign problem is comprehensively studied in a recent work~\cite{2021QPT}. The authors suggested a close relationship between the sign and the phase transition based on massive simulations of several models. This paper further explains how and when to use the sign to probe a phase transition. The general analysis independent of models or algorithms and examples in our work prove that the sign can only related to the critical points under special circumstances. The sign-free reference system must meet strict requirements to extract meaningful and non-misleading information on the phase transition from the sign. 
%For example, the partition function should vary more slowly than the original system, or the distance of the partition functions between the system and reference system should be close enough. Our discussions cannot cover all possible situations in practical calculations, and there can be more cases where the sign can successfully catch the phase transition. However, in general, searching for a proper reference system and making use of the sign is not easier than neutralizing the sign problem itself. 
Enlightened by those analyses, we propose a modified sign and put forward an algorithm calculating it with high precision, which can be used to calculate the deviation of free energy exactly in systems with sign problems. We believe that this new quantity would help probing the phase transitions in sign-problem systems.

\section{Acknowledgments}--We acknowledge Xu Zhang, Zi Yang Meng, Kai Sun for useful discussions. This research was supported by the National Natural Science Foundation of China (grant nos. 12004020, 12174167, 12247101), and the Fundamental Research Funds for the Central Universities in China.
The authors acknowledge Beijng PARATERA Tech CO.,Ltd. for providing HPC resources that have contributed to the research results reported within this paper. ZY thanks the start-up fund of the Westlake University.

\bibliography{sign}

\end{document}